\documentclass[fleqn,usenatbib]{mnras}

\usepackage{newtxtext,newtxmath}
\usepackage[T1]{fontenc}

\DeclareRobustCommand{\VAN}[3]{#2}
\let\VANthebibliography\thebibliography
\def\thebibliography{\DeclareRobustCommand{\VAN}[3]{##3}\VANthebibliography}

\usepackage{graphicx}	% Including figure files
\usepackage{amsmath}	% Advanced maths commands
\usepackage[dvipsnames]{xcolor} % Used for highlighting revisions
\usepackage{booktabs} % for nice tables
\title[Slow Alfv\'enic Coronal Hole Wind at 0.13 au]{Plasma Properties, Switchback Patches and Low $\alpha$-Particle Abundance in Slow Alfv\'enic Coronal Hole Wind at 0.13 au}

\author[T. Woolley et al.]{Thomas Woolley,$^{1}$\thanks{E-mail: thomas.woolley15@imperial.ac.uk}
Lorenzo Matteini,$^{1}$
Michael D. McManus, $^{2}$
Laura Berčič,$^{3}$
Samuel T. Badman,$^{2,4}$
\newauthor
Lloyd D. Woodham,$^{1}$
Timothy S. Horbury,$^{1}$
Stuart D. Bale,$^{2,4}$
Ronan Laker,$^{1}$
Julia E. Stawarz,$^{1}$
\newauthor
Davin E. Larson,$^{4}$
\\
$^{1}$ Department of Physics, Imperial College London, London SW7 2AZ, UK \\
$^{2}$ Physics Department, University of California, Berkeley, CA 94720-7300, USA \\
$^{3}$ Mullard Space Science Laboratory, University College London, Dorking, RH5 6NT, UK \\
$^{4}$ Space Sciences Laboratory, University of California, Berkeley, CA 94720-7450, USA}

\date{Accepted XXX. Received YYY; in original form ZZZ}

\pubyear{2021}

\begin{document}
\label{firstpage}
\pagerange{\pageref{firstpage}--\pageref{lastpage}}
\maketitle

\begin{abstract}
The Parker Solar Probe (PSP) mission presents a unique opportunity to study the near-Sun solar wind closer than any previous spacecraft. During its fourth and fifth solar encounters, PSP had the same orbital trajectory, meaning that solar wind was measured at the same latitudes and radial distances. We identify two streams measured at the same heliocentric distance ($\sim$0.13au) and latitude ($\sim$-3.5$^{\circ}$) across these encounters to reduce spatial evolution effects. By comparing the plasma of each stream, we confirm that they are not dominated by variable transient events, despite PSP's proximity to the heliospheric current sheet. Both streams are consistent with a previous slow Alfv\'enic solar wind study once radial effects are considered, and appear to originate at the Southern polar coronal hole boundary. We also show that the switchback properties are not distinctly different between these two streams. Low $\alpha$-particle abundance ($\sim$ 0.6 \%) is observed in the encounter 5 stream, suggesting that some physical mechanism must act on coronal hole boundary wind to cause $\alpha$-particle depletion. Possible explanations for our observations are discussed, but it remains unclear whether the depletion occurs during the release or the acceleration of the wind. Using a flux tube argument, we note that an $\alpha$-particle abundance of $\sim$ 0.6 \% in this low velocity wind could correspond to an abundance of $\sim$ 0.9 \% at 1 au. Finally, as the two streams roughly correspond to the spatial extent of a switchback patch, we suggest that patches are distinct features of coronal hole wind.
\end{abstract}

\begin{keywords}
Sun: heliosphere - solar wind - magnetic fields
\end{keywords}

%%%%%%%%%%%%%%%%%%%%%%%%%%%%%%%%%%%%%%%%%%%%%%%%%%

%%%%%%%%%%%%%%%%% BODY OF PAPER %%%%%%%%%%%%%%%%%%

\section{Introduction}
The solar wind, which is usually categorised as either fast ($>600$km/s) or slow ($<400$km/s) based on its bulk velocity, comprises ions and electrons that stream out into interplanetary space. Protons are by far the most dominant ion species, but the plasma also contains populations of doubly-ionised helium ($\alpha$-particles) and heavier ions. It is widely accepted that the source of the fast solar wind is coronal holes \citep{Krieger_1973, Nolte_1976, Sheeley_1976, Neugebauer_1998}, but the origin of the more variable slow solar wind \citep{Bame_1977} is still unclear. Slow solar wind could come from coronal hole boundaries at solar minimum \citep{wang_sheeley_1994, Neugebauer_1998}, and from small coronal holes \citep{wang_sheeley_1994, Neugebauer_1998} and/or active regions at solar maximum \citep{Neugebauer_2002, Harra_2008}.

Part of the issue with identifying the sources of the slow wind arises because of this binary classification based purely on bulk velocity. Many studies have therefore tried to improve upon the current solar wind categorisation \citep[e.g.][]{xu_borovsky_2015, Camporeale_2017, Li_2020}. Recent work by \cite{Damicis_Bruno_2015} suggests that there are at least two modes of slow solar wind: an Alfv\'enic population and a non-Alfv\'enic population. The Alfv\'enic slow wind has similar properties to the fast wind \citep{Damicis_2019, Perrone_2020, stansby_2020}, which indicates that it may also originate in coronal holes or at their boundaries \citep{Damicis_Bruno_2015}. The non-Alfv\'enic slow wind is associated with highly variable mass fluxes \citep{Stansby_2019}, which is consistent with wind from active regions \citep{Damicis_Bruno_2015} or transients related to coronal streamers \citep{Einaudi_1999, Damicis_Bruno_2015, Stansby_2019}. However, more studies are needed to fully characterise different solar wind streams as the sources and formation mechanisms of slow wind are still actively debated \citep{Abbo_2016}. The recent launch of the Parker Solar Probe (PSP; \citet{Fox_2016}) mission provides a novel data set for these studies.

To date, PSP has remained at low latitudes during solar minimum and predominantly observed slow solar wind. PSP has routinely measured switchbacks - rotations of the magnetic field away from its background orientation - in this slow wind \citep{Bale_Nature_2019, Kasper_nature_2019}, but switchbacks have also been previously observed in fast solar wind by Helios \citep{Horbury_2018} and Ulysses \citep{Neugebauer_2013}. Switchbacks appear to occur in groups or `patches' which are separated by more radial `quiet' intervals close to the Sun \citep{Horbury_2020}. While many studies have looked at the properties of switchbacks including their shape and size \citep{Laker_2020}, their temperature \citep{Woolley_2020, Woodham_2020} and their occurrence rate and duration \citep{DudokdeWit_2020}, the mechanisms that cause switchbacks are still actively debated. It has been suggested that magnetic field switchbacks could be related to solar wind origins. For example, switchbacks may be caused by interchange reconnection \citep{Zank_2020}, mini-filament eruption \citep{Sterling_2020} or Alfv\'en-wave turbulence \citep{Squire_2020, Shoda_2021}, meaning that studies of these events could help to disentangle the physical processes governing the solar wind release.

The $\alpha$-particle (or helium) abundance ($A_{\alpha}$) - the ratio of $\alpha$-particle to proton number density - is another feature that is highly influenced by solar source regions. On average, the $\alpha$-particle abundance in the solar wind is between 3\% and 6\% \citep{Ogilvie_1969, Robbins_1970, Neugebauer_1981, Alterman_2021}, but can increase to over 15\% during solar flares \citep{Hirshberg_1972b}. At solar minimum, the $\alpha$-particle abundance is higher \citep{Hirshberg_1972, Kasper_2007, Kasper_2012} and more stable \citep{Bame_1977, feldman_1978} in the fast solar wind ($A_{\alpha} \sim$ 4-5\%) than in the slow wind ($A_{\alpha} \sim$ 1-2\%). However, this velocity dependence does not persist during solar maximum, where the slow and fast wind have comparable $\alpha$-particle abundances of $\sim$ 4-5\% \citep{Kasper_2007, Kasper_2012}. Recent work by \citet{Huang_2020_b} shows that highly Alfv\'enic slow solar wind contains both helium poor ($A_{\alpha} \sim$ 1.5 \%) and helium rich ($A_{\alpha} \sim$ 4.5 \%) populations, suggesting multiple solar sources for this type of wind. They also show that some slow solar wind plasma properties are similar regardless of Alfv\'enicity, which could indicate commonality in the source regions of all slow solar wind.

A crucial step to understanding the sources and release of solar wind plasma involves identifying the properties of different solar wind types closer to the Sun. Here we take advantage of the fact that PSP had the same orbital trajectory during encounters 4 and 5 to identify two streams of slow Alfv\'enic coronal hole wind measured at the same latitude and radius during solar minimum. We compare the proton and $\alpha$-particle parameters before presenting the conditions of this slow Alfv\'enic wind closer than any previous study. In Section \ref{switchback_statistics}, we analyse switchback statistics for each interval.  We present $\alpha$-particle data for one stream, showing that the $\alpha$-particle abundance is very low and justifying why such low abundance is not seen at 1 au. These results are then discussed in relation to solar wind sources in Section \ref{low_alpha_discussion}.

\section{Data}

The data used here were measured by PSP \citep{Fox_2016} during its fourth and fifth encounters. The magnetic field data were measured by the flux-gate magnetometer (MAG) from the FIELDS instrument suite \citep{Bale_2016}.

We use proton core parameters (i.e. velocity, parallel temperature, perpendicular temperature and density) derived from the ion distribution functions measured by the SPAN-I instrument \citep{Kasper_2016}. We fit a two population bi-Maxwellian to the 3D proton distribution functions as detailed in \citet{Finley_2020} \citep[For alternatives see e.g.][]{Stansby_2018_helios_fits, Woodham_2020}. In the presence of Alfv\'enic fluctuations, the proton core and beam populations rotate in velocity space \citep{Matteini_2014, Matteini_2015_alfvenic} and can move out of the instrument's limited field of view \citep{Woodham_2020}, making measurements of these populations unreliable under certain plasma conditions. For the intervals studied here we focus only on the proton core as this is more steadily within the field of view than the proton beam. The $\alpha$-particles, which are approximately stationary in velocity space under Alfv\'enic fluctuations \citep{Matteini_2015_alfvenic}, were fit using a bi-Maxwellian after removing proton contamination from the $\alpha$-particle channel as in \citet{Finley_2020}. The encounter 4 $\alpha$-particle data have been disregarded here because unusually low counts in the instrument resulted in less reliable fits.

We use strahl electron parallel temperature ($T_{strahl,\parallel}$) as an estimate for the coronal electron temperature at the source \citep{Bercic_2020}. To obtain this parameter we fit the electron velocity distribution functions (VDFs) measured by SPAN-E instrument \citep{Kasper_2016, Whittlesey_2020}. We perform the fits in the magnetic field aligned, plasma rest frame, which is defined with the magnetic field vector from MAG and the velocity moment derived from the SPAN-I proton distribution functions. We model the parallel cut through the strahl distribution function with a non-drifting 1D Maxwellian, from which we obtain $T_{strahl,\parallel}$. According to the collisionless exospheric models \citep{Jockers_1970, Lemaire_1970, Lemaire_1971, Maksimovic_1997} this temperature does not change during the solar wind expansion and can therefore give a good estimate of the electron temperature in the solar corona. The fitting method includes an additional step to correct for the portion of the SPAN-E field-of-view blocked by the PSP’s heat shield, which is further described in Sec. 3.1 of \citet{Bercic_2020}.

\begin{figure*}
\centering
 \includegraphics[width=2.0\columnwidth]{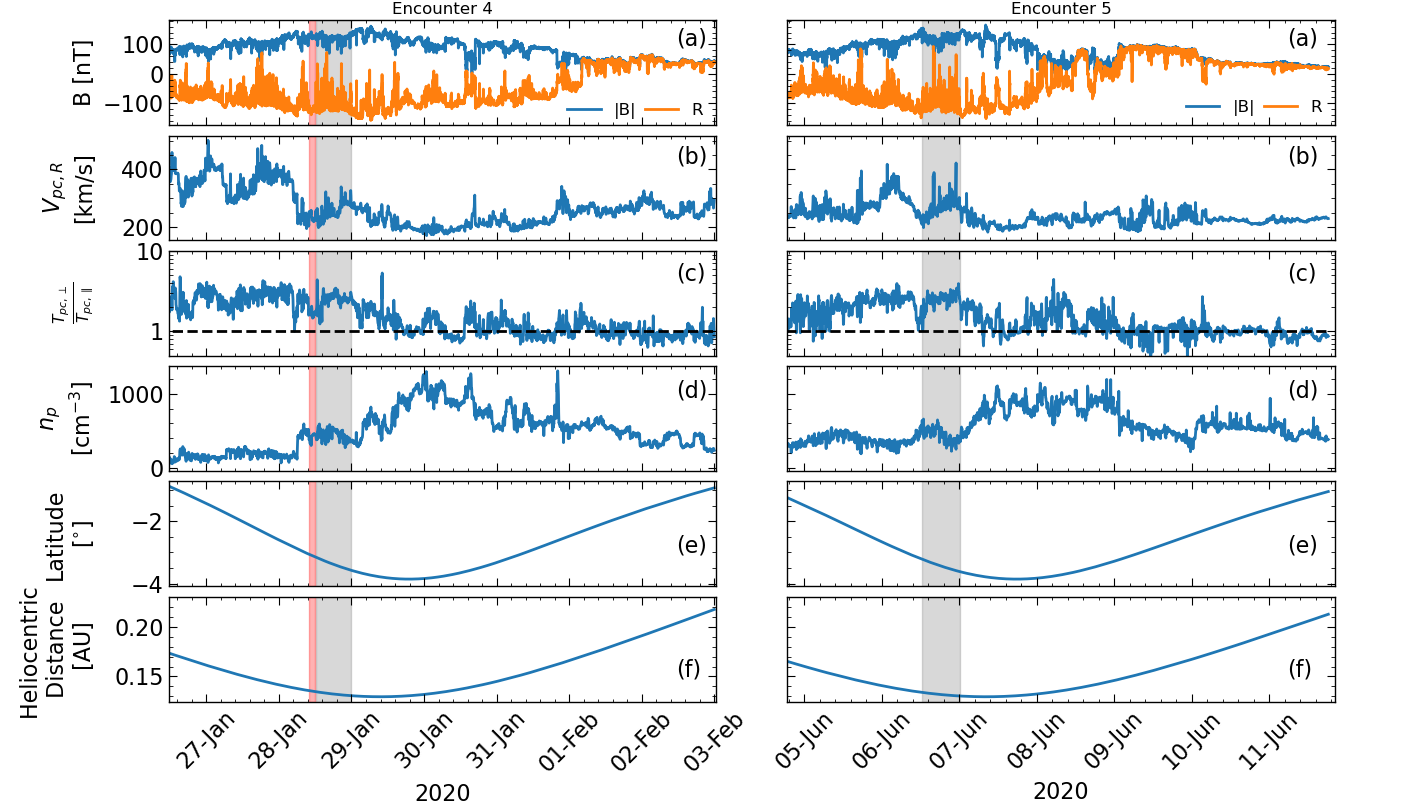}
\caption[]{\small Time series of PSP's encounter 4 (left) and encounter 5 (right). The panels show: (a) magnitude and radial component of the magnetic field, (b) proton core radial velocity, (c) proton core temperature anisotropy, (d) total proton number density, (e) latitude and (f) heliocentric distance. The grey shaded regions show the two streams analysed in this study. The red shaded region shows the `quiet' interval used for comparison in this study. The dashed black line in panel (c) shows where the temperature anisotropy is unity.}
\label{fig1_timeseries}
\end{figure*}

\section{Results}

\subsection{Identification of Two Streams}
\label{stream_identification}
In order to reduce spatial effects, we manually choose two streams across PSP's fourth and fifth encounters such that the solar wind was measured at the same latitude and radius in each. Fig. \ref{fig1_timeseries} shows time series data from encounters 4 (left) and 5 (right), with the two intervals used in this study shaded grey (referred to now on as the E4 and E5 streams, respectively). Panel (a) shows the magnitude and the radial component of the magnetic field, which appears to be modulated into `patches' of switchbacks separated by `quiet' solar wind \citep{Bale_Nature_2019, Horbury_2020}. This `quiet' solar wind (referred to as quiet intervals or periods henceforth) has been shown to contain large proton beams and significant wave activity \citep{Verniero_2020}. The red shaded region is a `quiet' period and is used as a comparison interval to the two streams identified. Due to the high Alfv\'enicity of the solar wind close to the Sun, the proton core radial velocity, which is $\sim$ 300 km/s in the two streams, also appears in patches (panel b). Panels (c) and (d) show the proton core anisotropy and the proton density, respectively. During PSP's outbound journey for each encounter, the density is enhanced and the anisotropy reduced compared to the same radius on the inbound journey. This is because of the proximity of PSP to the heliospheric current sheet (HCS) and the heliospheric plasma sheet (HPS), where slower and more dense wind is found. Panels (e) and (f) show that both streams were measured at $\sim$-3.5$^{\circ}$ latitude and $\sim$0.13 au, just before perihelion.

\subsection{Comparison of the Streams}
\label{stream_comparison}

\subsubsection{Plasma Properties}
\label{stream_plasma_properties}
\begin{figure*}
\centering
 \includegraphics[width=2.0\columnwidth]{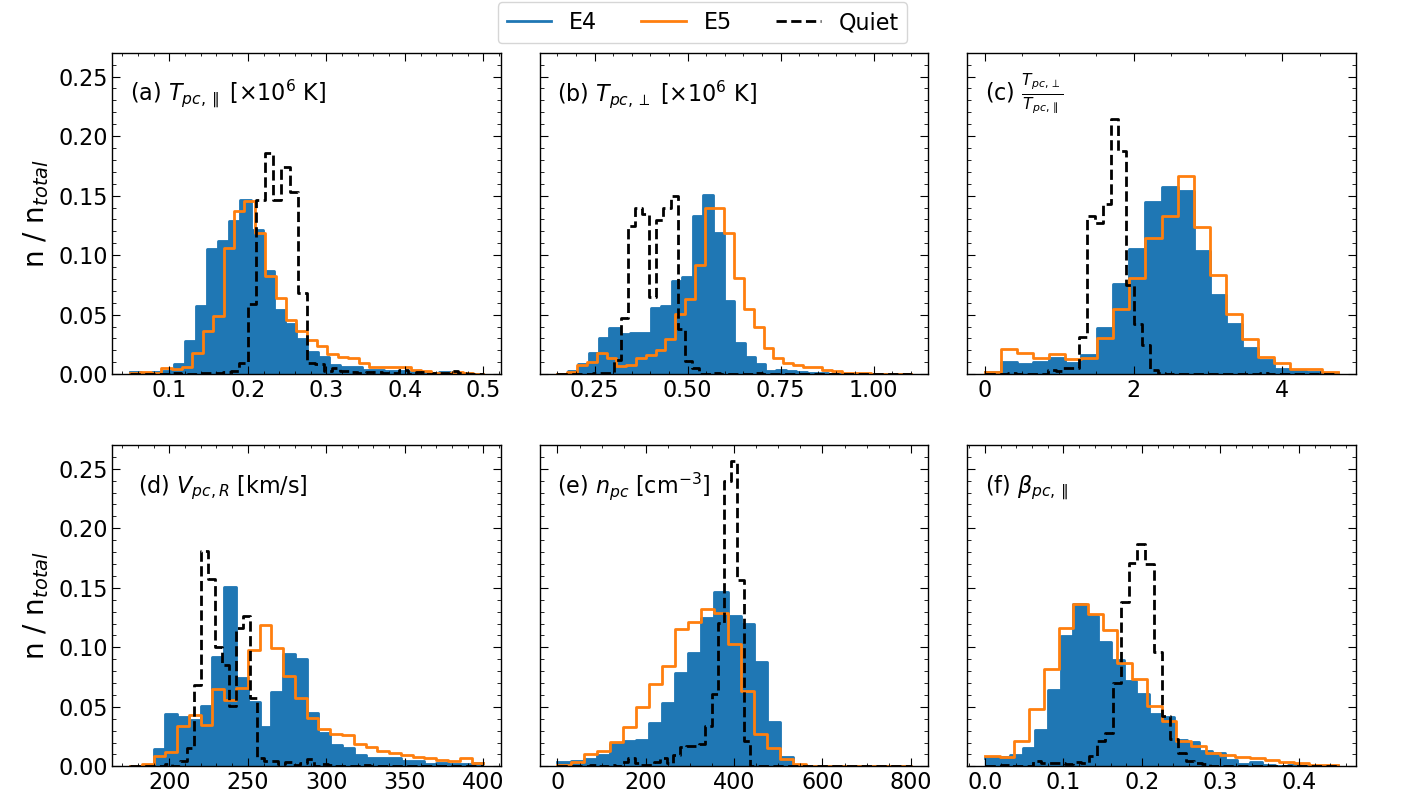}
\caption[]{\small Histograms of proton core parameters during the E4 (blue), E5 (orange) and quiet (black) streams. The panels show: (a) parallel temperature, (b) perpendicular temperature, (c) temperature anisotropy, (d) radial velocity, (e) number density and (f) plasma $\beta_{\parallel}$.}
\label{fig2_proton_core}
\end{figure*}

Fig. \ref{fig2_proton_core} shows distributions of proton core parameters for the two streams identified in Section \ref{stream_identification}. The panels show: (a) parallel temperature, (b) perpendicular temperature, (c) temperature anisotropy, (d) radial velocity, (e) number density and (f) plasma $\beta_{\parallel}$. The neighbouring interval of `quiet' wind (see red shaded interval in Fig. \ref{fig1_timeseries}) measured at a similar radius, latitude and time is also shown here.

Overall, the proton core distributions in the E4 and E5 streams are remarkably similar to each other for solar wind measured 5 months apart. The perpendicular temperature distributions show the largest difference with the E5 stream $\sim 13\%$ hotter on average than the E4 stream. This is most likely caused by the solar wind temperature-velocity relationship \citep{Burlaga_1970, Lopez_1986}, as the E5 stream is, on average, $\sim 5\%$ faster than the E4 stream. This velocity difference could also explain the slightly lower density in the E5 stream as faster wind is less dense than slower wind \citep{Verscharen_2019}. In contrast, the parallel temperature distributions are very similar and peak at $\sim 0.2 \times 10^{6}$K. This could indicate that parallel temperature is less strongly dependent on velocity than the perpendicular temperature, as previously predicted for fast solar wind at 35$R_{S}$ \citep{Perrone_2019_thermo}. Finally, the temperature anisotropy and $\beta_{pc, \parallel}$ distributions show no significant differences and peak at $\sim2.5$ and $\sim0.1$ for both streams. Some of the quiet interval parameters deviate significantly from the E4 and E5 streams. For example, the perpendicular temperature is considerably lower and the parallel temperature is higher, leading to a smaller anisotropy. The `quiet' period radial velocity and density distributions are similar to, but more strongly peaked than, the corresponding E4 and E5 stream distributions. $\beta_{pc, \parallel}$ in the `quiet' interval is almost twice as large as in the E4 and E5 streams.

\begin{figure*}
\centering
 \includegraphics[width=2.0\columnwidth]{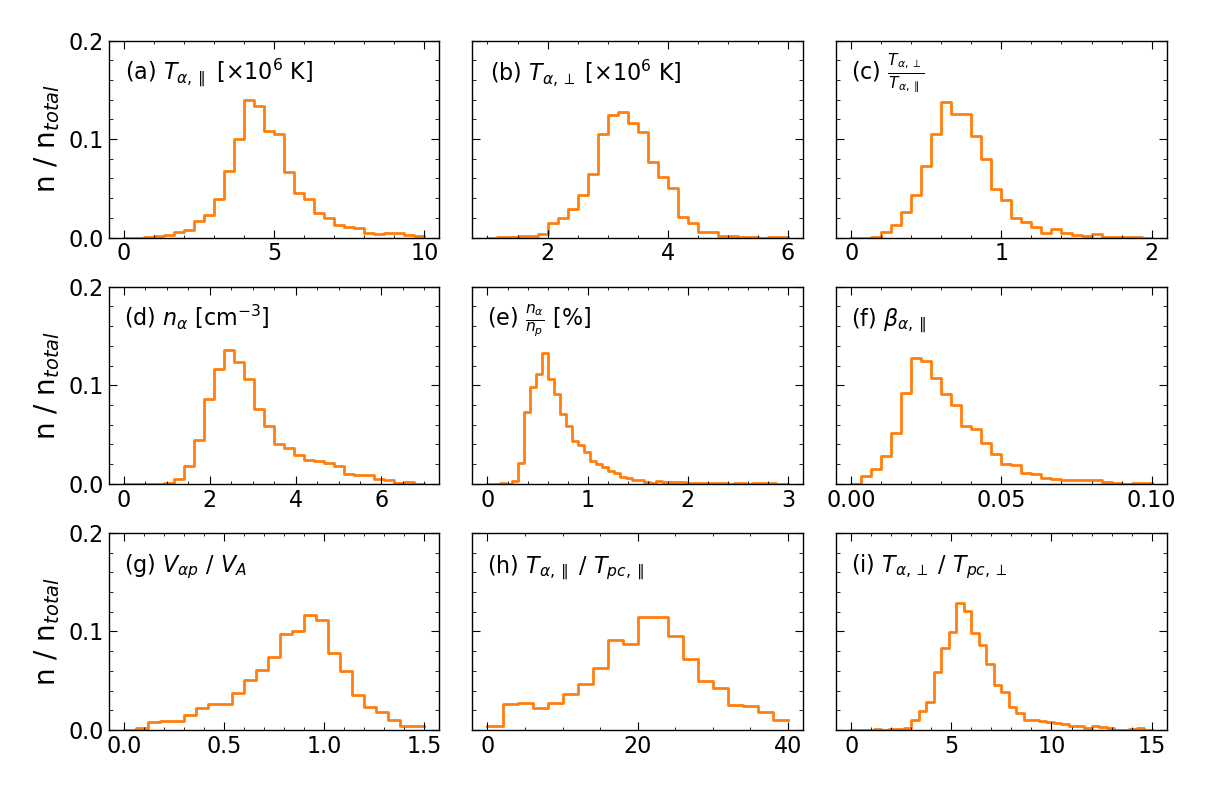}
\caption[]{\small Histograms of $\alpha$-particle parameters during the E5 stream. The panels show: (a) parallel temperature, (b) perpendicular temperature, (c) temperature anisotropy, (d) number density, (e) $\alpha$-particle abundance, (f) plasma $\beta_{\parallel}$, (g) normalised $\alpha$-proton drift, (h) $\alpha$-proton parallel temperature ratio and (i) $\alpha$-proton perpendicular temperature ratio.}
\label{fig3_alphas}
\end{figure*}

Fig. \ref{fig3_alphas} shows distributions of the E5 stream $\alpha$-particle parameters. Panels (a) and (b) show that the $\alpha$-particles have parallel and perpendicular temperatures of $\sim$ 4.5$\times$10$^{6}$K and $\sim$ 3.5$\times$10$^{6}$K, respectively. The $\alpha$-particles are considerably hotter than the protons here. The $\alpha$-to-proton temperature ratios are $\sim$ 20-25 and $\sim$ 6 for parallel and perpendicular, respectively (panels (h) and (i)). Such a large parallel ratio may seem nonphysical, but it is a consequence of protons and $\alpha$-particles having opposite anisotropy.  Panel (c) shows that the $\alpha$-particles, in contrast to the protons, have a temperature anisotropy less than one. The density of $\alpha$-particles is $\sim$ 3cm$^{-3}$ (panel d), which gives a remarkably low abundance (panel e). This is discussed in detail in Section \ref{low_alpha_discussion}. Panel (f) shows $\beta_{\alpha, \parallel}$ is $\sim$ 0.025, an order of magnitude less than protons. Panel (g) shows that the $\alpha$-particles, like in coronal hole fast wind, drift at a large fraction of the Alfv\'en speed ahead of the proton core population.

\subsection{Type of Solar Wind and Source Region}
\label{types_of_sw}
As our comparisons suggest that the solar wind in the E4 and E5 streams is quantitatively similar, we conclude that this wind is unlikely to be caused by transient events which tend to be much more variable. We therefore try to establish the solar source of these solar wind streams.

Power spectra of the magnetic field fluctuations for each stream show power laws of -3/2 in the inertial range (not shown here). This is consistent with the scaling of non-HPS solar wind previously shown by \citet{Chen_2021} using a similar encounter 4 stream. The -3/2 power law is also consistent with solar wind from the small equatorial coronal hole in encounter 1 \citep{Bale_Nature_2019, Chen_2020}. Conversely, the wind measured during the outbound sections of the encounters has an inertial range power law of -5/3 \citep{Chen_2021}. The wind during this time was slower and more dense, consistent with predominantly HPS wind. We conclude that the chosen streams, measured on the inbound section of each encounter, are non-HPS solar wind from a coronal hole.

Further evidence to support our conclusion is obtained by estimating the solar source region. The two solar wind streams are ballistically mapped to the source surface before a potential field source surface (PFSS) model \citep{Altschuler_1969, Schatten_1969} is applied, using the \texttt{pfsspy} library \citep{Stansby2020_pfsspy}, to determine solar origins of the wind \citep[For other studies that use PFSS mapping see e.g.][]{Badman_2020, Laker_2021, Kane_2021}. Both streams originate at the boundaries of the Southern polar coronal hole, separated by $\sim$30$^{\circ}$ in longitude, according to the model (See Fig. \ref{pfss_map} in Appendix. \ref{pfss_appendix}). While this mapping is consistent with the in situ measured polarity, PSP is located very close to the HCS in both streams and so additional evidence to support the mapping and rule out streamer belt contributions is desirable. Considering this mapping with the high $\alpha$-proton drift (See Fig. \ref{fig3_alphas}) - a property of fast solar wind \citep{Berger_2011, Alterman_2018} which is often linked to coronal hole source regions - further suggests that this slow wind originates from the boundaries of the Southern polar coronal hole.

Finally, Fig. \ref{electron_strahl} shows the strahl electron parallel temperature for each stream. As the strahl electrons propagate out of the corona with very little interaction with thermal distributions, this can be used as a proxy for the source temperature \citep{Bercic_2020}. For the two streams, the strahl parallel temperature exhibits statistically similar distributions that peak around 1.1 $\times$10$^{6}$K, suggesting that the two streams originated in regions with similar local plasma temperatures. By combining the above results with the low velocity of the wind, we suggest that the two streams contain slow Alfv\'enic coronal hole boundary wind from similar temperature source regions.

Table \ref{table_stansby} shows a comparison of the slow Alfv\'enic wind at 0.13 au and at 0.35 au \citep{stansby_2020}. The largest difference between the two studies is seen in the radial velocity, with the wind at 0.13 au being $\sim$ 100 km/s slower. There is also a larger proton anisotropy at 0.13 au, which is due to the radial evolution of temperature with distance from the Sun. The normalised $\alpha$-proton drift is larger closer to the Sun, while the proton flux and $\alpha$-particle anisotropy are similar at the two radial distances. The $\alpha$-proton temperature ratios are not markedly different in the two studies. These results are discussed further in Section \ref{discussion}.

\begin{table}
\begin{tabular}{@{}lcc@{}}
\toprule
Quantity & \multicolumn{1}{l}{This Study} & \multicolumn{1}{l}{Stansby et al. (2020b)} \\ \midrule
Radius {[}au{]}               & 0.13        &  0.35 \vspace{0.1cm} \\
$V_{pc, R}$ {[}km/s{]}         & $\sim$250 - 270  &  $\sim$380 \vspace{0.1cm}   \\
$n_{p}V_{pc, R}r^{2}$ {[}$10^{35}s^{-1}sr^{-1}${]}  & $\sim$0.5  &  $\sim$0.6  \vspace{0.1cm} \\
%Beam Fraction {[}\%{]} &  $\sim$19-25       &  $\sim$17.5    \vspace{0.1cm}  \\
Proton Anisotropy             & $\sim$2-3   &  $\sim$1-2     \vspace{0.1cm}  \\
$\alpha$-proton $T_{\perp}$ ratio    &  $\sim$6    &    $\sim$4  \vspace{0.1cm}  \\
$\alpha$-proton $T_{\parallel}$ ratio &  $\sim$20-25   & $\sim$20 \vspace{0.1cm} \\
Normalised $\alpha$-proton Drift        &  $\sim$0.9  & $\sim$0.5             \vspace{0.1cm}  \\
$\alpha$ Anisotropy             &  $\sim$0.5-0.6    &  $\sim$0.6-0.7            \vspace{0.1cm}  \\ \bottomrule
\end{tabular}
\caption[]{\small Peak distribution values of slow Alfv\'enic wind
parameters from this study and \citet{stansby_2020}.}
\label{table_stansby}
\end{table}

\subsubsection{Switchbacks}
\label{switchback_statistics}
Table \ref{table_switchbacks} summarises the statistical properties of the switchbacks in each stream (the quiet interval has not been included as, by definition, quiet intervals have very few switchbacks). Switchbacks were identified using 4 Hz magnetic field data. A switchback was defined as when the magnetic field cone angle - the angle between the magnetic field vector and the radial direction - deflected by more than 20$^{\circ}$ from the 5 hour background average. Any switchbacks with a duration of less than 1s (4 data points) were discarded. It should be noted that there is currently no consensus for consistently identifying switchbacks, and so methods differ between studies.

Overall, the switchback statistics between the two streams are very similar. In particular, the number of switchbacks in the E4 stream (817) and in the E5 stream (851) are not markedly different. Approximately 16\% of each stream is part of a switchback and the waiting times - the time between one switchback ending and the next starting - follow power laws with -1.44$\pm$ 0.04 and -1.51$\pm$ 0.10 dependencies for the E4 and E5 streams, respectively. These are consistent with the power laws found in \citet{DudokdeWit_2020}. The tendency for larger deflection switchbacks to have longer durations \citep{Horbury_2020} is also clear in both streams, with correlation coefficients between average deflection and duration of 0.58 (E4 stream) and 0.69 (E5 stream).

The threshold of 20$^{\circ}$ was arbitrarily chosen as there is no definite way of identifying switchbacks. Increasing this threshold reduces the number of switchbacks and the proportion of the stream that comprises switchbacks, but it does not significantly affect the correlation between average deflection angle and duration. The waiting time power laws appear to vary between -1.3 for higher thresholds (i.e. 40$^{\circ}$) and -1.6 for lower thresholds (i.e. 5$^{\circ}$).

\begin{table}
\begin{tabular}{@{}lcc@{}}
\toprule
Quantity  & \multicolumn{1}{l}{E4 Stream} & \multicolumn{1}{l}{E5 Stream} \\ \midrule
\# of SBs              & 817    & 851       \vspace{0.15cm}   \\
SB \% of Stream         & 16.4     & 15.6     \vspace{0.15cm}     \\
Deflection-Duration Correlation  & 0.58    & 0.69   \vspace{0.15cm}  \\
Waiting Time Power Law     & -1.44 $\pm$ 0.04   & -1.51 $\pm$ 0.10  \vspace{0.15cm}  \\ \bottomrule
\end{tabular}
\caption[]{\small Switchback (SB) statistics from the E4 and E5 streams.}
\label{table_switchbacks}
\end{table}

\section{Discussion}
\label{discussion}
In Section \ref{types_of_sw}, we concluded that the E4 and E5 streams contained slow Alfv\'enic coronal hole wind. We can therefore discuss the results in this context.

\subsection{Plasma Properties}

Fig. \ref{fig1_timeseries} and \ref{fig2_proton_core} show that the solar wind radial velocity of each stream is less than 300 km/s and therefore much lower than the typical speed of wind measured at 1 au. In fact, solar wind with velocities $\leq$ 300 km/s is rarely seen at 1 au. This could be a consequence of stream interactions, which increase the velocity of slower wind and decrease that of faster wind \citep{Gosling_1978, Breen_2002}. While the two streams have low velocities, they appear to be a different type of solar wind to the very slow solar wind presented in \citet{Sanchez_Diaz_2016}.

Fig. \ref{fig1_timeseries} shows that the proton core anisotropy of the slow Alfv\'enic wind at 0.13 au is larger than at 0.35au (See Table \ref{table_stansby}). This is consistent with $T_{pc, \perp}$ decreasing faster with increasing radius than $T_{pc, \parallel}$, as previously shown for fast solar wind \citep{Marsch_1982_protons, Hellinger_2011}. This also suggests that the protons are preferentially heated in the perpendicular direction closer to the Sun by a mechanism such as stochastic heating \citep{Chandran_2010, Bourouaine_2013}, ion-cyclotron resonance \citep{Li_1999}, or dissipation of turbulence \citep{Smith_2001, Verdini_2010}. The $\alpha$-particle temperature anisotropy, however, is comparable at 0.13 au and 0.35 au (See Table \ref{table_stansby}), suggesting that it does not evolve significantly between these two distances. As $\alpha$-particles have been shown to drive similar plasma instabilities to protons \citep{Maruca_2012}, such as the firehose instability \citep{Matteini_2015}, it is possible that the evolution of $\alpha$-particles has already taken place by 0.13 au and hence they remain bounded by instability thresholds. The $\alpha$-particle anisotropy is in the opposite sense to protons (i.e. $T_{\alpha, \perp} < T_{\alpha, \parallel}$), which is consistent with previous fast wind studies \citep{Marsch_1982, Stansby_2019_alphas}. This could be due to the earlier evolution of $\alpha$-particles or a signature that different heating mechanisms act on different ion species. In contrast, there is a clear reduction in the $\alpha$-proton drift between the two studies, which could be related to instabilities \citep{Matteini_2012, Hellinger_2013} or expansion \citep{Verscharen_2015}, and the redistribution of this drifting kinetic energy could act to maintain an $\alpha$-particle anisotropy close to unity.

\subsection{Low \texorpdfstring{$\alpha$-particle Abundance}{}}
\label{low_alpha_discussion}

An interesting result presented in Section \ref{stream_plasma_properties} is the very low $\alpha$-particle abundance ($\sim$0.6\%) found during the E5 stream. This is much lower than the average abundance seen at 1 au, which can be partly explained by a simple flux tube argument as presented by \citet{Kasper_2007}. By considering the motion of protons and $\alpha$-particles in a one-dimensional flux tube, \citet{Kasper_2007} showed that the $\alpha$-particle abundance would increase as the relative drift between protons and $\alpha$-particles decreases. As $\alpha$-particles drift at a large fraction of the Alfv\'en speed ahead of the proton core in the slow Alfv\'enic solar wind (See Fig. \ref{fig3_alphas}), and the Alfv\'en speed drops with radial distance, the $\alpha$-particle abundance increases with radial distance. Using typical values for the solar wind and Alfv\'en speed, we find that an $\alpha$-particle abundance of $\sim$ 0.6\% at 0.13 au would lead to an abundance of $\sim$  0.9\% at 1 au, which is comparable to previous observations \citep{Alterman_2021}. However, this value of $\alpha$-particle abundance is still small when compared to faster coronal hole wind, and therefore other mechanisms are needed to explain the range of $\alpha$-particle abundances seen in the solar wind.

Ideas of $\alpha$-particle variability are closely linked to solar wind release and heating mechanisms. There are two competing theories surrounding the source of the solar wind and what controls its properties. The first theory suggests that plasma confined to closed loops in the corona is released by interchange reconnection with open field lines \citep{Fisk_1999, Fisk_2001}. It is the properties of the closed loop plasma that determine the final properties of the wind in this scenario \citep{Fisk_1999}. For example, it has been shown that the solar wind speed squared is inversely proportional to the loop temperature \citep{Fisk_2003, Gloeckler_2003} and that hotter loops are typically larger \citep{Feldman_1999}. \citet{Rakowski_2012} showed that in larger loops the First Ionisation Potential (FIP) effect \citep{Laming_2004} causes more helium depletion than in shorter loops, and argued that gravitational settling \citep{Byhring_2011} is too slow to cause the variability seen. Combining these results suggests that slow solar wind originates from reconnection between open field and large, hot loops with depleted amounts of $\alpha$-particles.

The second theory suggests that plasma flows outwards on open field lines and requires no interchange reconnection to determine the properties of the wind. Instead, the solar wind properties are determined by the amount of expansion that flux tubes of solar wind undergo. For example, \citet{Cranmer_2007} showed using coronal heating models that varying only the magnetic field could reproduce realistic solar wind values and correlations. \citet{Geiss_1970} suggested that a minimum proton flux is required to accelerate helium into the solar wind via dynamical friction, and if the proton flux is too low, then the proportion of helium can be depleted. However, \citet{stansby_2020} showed there was no obvious dependence of $\alpha$-particle abundance on proton flux between fast and slow wind streams, concluding that other mechanisms must also be responsible.

From our study it is not clear which of these theories is most consistent with PSP observations, but recent work by \citet{Fu_2018} suggests that these two mechanisms can work in parallel, with the open field line mechanism being more important within coronal holes. A more complicated framework for solar wind release and acceleration has been proposed by \citet{Viall_2020_review} to address observed cases that are not easily resolved by these theories. This framework includes both interchange reconnection and open field lines as mechanisms for the solar wind to leave the corona, but assumes that each source can, under different conditions, be governed by multiple release and acceleration processes. However, further work using a combination of in-situ and remote sensing data is needed to bridge the gap between in situ measurements and these theories.

Finally, we can use the simple flux tube argument to make a prediction of the $\alpha$-particle abundance at smaller heliocentric distances. As the Alfv\'en speed continues to increase closer to the Sun, there will be a point in the slow wind, the Alfv\'en critical point, where $v_{SW} \approx V_{A}$. Here we expect the $\alpha$-particle abundance to decrease by a further 25\% in slow solar wind streams similar to those presented in this study. This means that slow solar wind with an $\alpha$-particle abundance $<$ 0.5 \% could be measured in PSP's future orbits.

\subsection{Switchbacks and Patches}

The switchback statistics shown in Section \ref{switchback_statistics} suggest that the source regions produce switchbacks at a similar rate and in a similar quantity, even though they are separated by about 30$^{\circ}$ in longitude and the data were measured $\sim$5 months apart. This result indicates that the process leading to the formation of switchbacks does not display significant longitudinal or temporal variation, and that switchbacks, regardless of where they are measured in the near-Sun solar wind, have very similar properties. However, further work is needed to confirm if and how switchback statistics vary between different sources and distances.

\begin{figure}
  \includegraphics[width=\columnwidth]{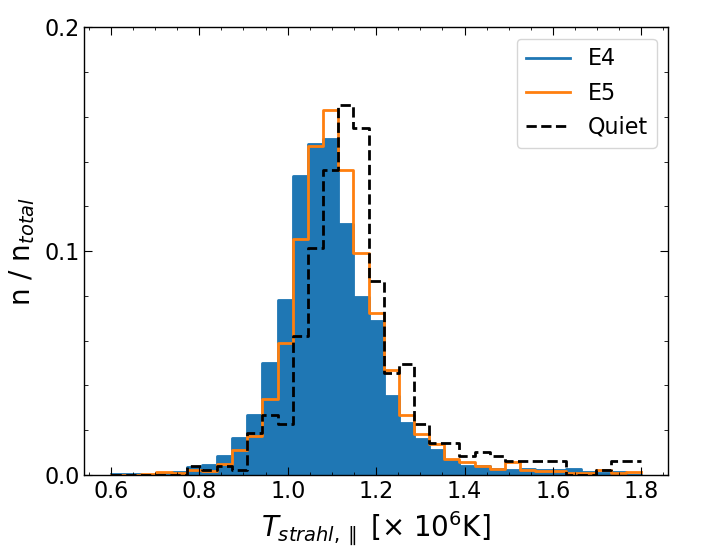}
 \caption[]{\small Distributions of the electron strahl temperatures during the E4 (blue), E5 (orange) and quiet (black) streams.}
 \label{electron_strahl}
 \end{figure}

The time series of the two streams we identified correspond approximately to the spatial extent of a patch of switchbacks \citep{Bale_Nature_2019, Horbury_2020}. As such, we suggest that the patches are distinct features of coronal hole wind. The quiet intervals could be spatial structure related to e.g. funnels \citep{Tu_2005} that PSP passes through, or transient events, such as streamer belt blobs \citep{Lavraud_2020}, that advect past the spacecraft and obscure the coronal hole wind from being measured. If the electron strahl temperature is a good measure of the source temperature, then Fig. \ref{electron_strahl} suggests that the quiet wind originates in regions with a similar temperature to the patch wind. This may indicate that quiet intervals also contain solar wind from coronal holes, however, a more in-depth treatment of patches and quiet periods is needed to determine this.

\section{Conclusion}

We have presented a comparison of two streams of slow Alfv\'enic wind measured at the same latitude and heliocentric distance by Parker Solar Probe (PSP) during its fourth and fifth solar encounters. From the statistically similar distributions of proton parameters (i.e. parallel and perpendicular temperatures, radial velocity, temperature anisotropy, density and $\beta_{\parallel}$) and electron strahl parallel temperature, we concluded that both wind streams had similar origins and were not associated with transient events.

We determined the source of this wind to be the boundary of the Southern polar coronal hole from PFSS modelling. The inertial range scaling of the power spectra and the high $\alpha$-proton drift velocities were consistent with coronal hole wind, and therefore were consistent with the solar source obtained from PFSS modelling. We explained with a flux tube argument from \citet{Kasper_2007} that the low $\alpha$-particle abundance seen at 0.13 au could reasonably increase by a factor of $\sim 1.5$ to $\sim$ 0.9\% at 1 au. This is consistent with previously measured $\alpha$-particle abundances in slow wind at 1 au \citep{Alterman_2021}. The flux tube argument however, is not sufficient to explain the variability in $\alpha$-particle abundance seen in the solar wind at all heliocentric distances and more advanced models are needed.

We suggested that because the switchback statistics did not significantly change between the two streams measured 5 months and $\sim$ 30$^{\circ}$ in surface longitude apart, the process leading to switchback formation does not display distinct temporal or spatial variation over these scales. Finally, as the two streams selected roughly corresponded to the spatial extent of switchback patches, we suggested that these patches are distinct features of coronal hole wind. The quiet periods that occur between patches could be related to spatial modulation in the solar source region or local transient events associated with the streamer belt. We proposed that, due to the similarity in the electron strahl temperature of the patches and the quiet period, that quiet periods could also be associated with coronal hole regions.

Future work could address how the $\alpha$-particle properties are linked to the solar source regions and characterise the `quiet' intervals and patches. These studies will benefit from the decreasing PSP orbital distance and the constellation of spacecraft, including Solar Orbiter \citep{Muller_2013} and Bepi Colombo, now taking in-situ measurements in the heliosphere.

\section*{Streams}

\begin{table}
\begin{tabular}{@{}lcc@{}}
\toprule
Stream       & Start Time & End Time \\ \midrule
E4 Stream    &    28 Jan 2020 12:07  & 28 Jan 2020 23:56  \\
E5 Stream    &    06 Jun 2020 12:23  & 07 Jun 2020 00:21  \\
Quiet Stream &    28 Jan 2020 10:00  & 28 Jan 2020 12:00  \\ \bottomrule
\end{tabular}
\caption[]{\small Start and end times for the E4, E5 and quiet streams.}
\end{table}

\section*{Data Availability}

The data used in this research is publicly available at: \url{https://cdaweb.gsfc.nasa.gov/index.html/}

\section*{Acknowledgements}
TW was supported by STFC grant ST/T506151/10. TSH and LDW by ST/S000364/1 and RL by an Imperial College President’s scholarship. JES was supported by the Royal Society University Research Fellowship URF/R1/201286.

\appendix

\section{Potential Field Source Surface Mapping}
\label{pfss_appendix}
\begin{figure*}
\centering
 \includegraphics[width=2.0\columnwidth]{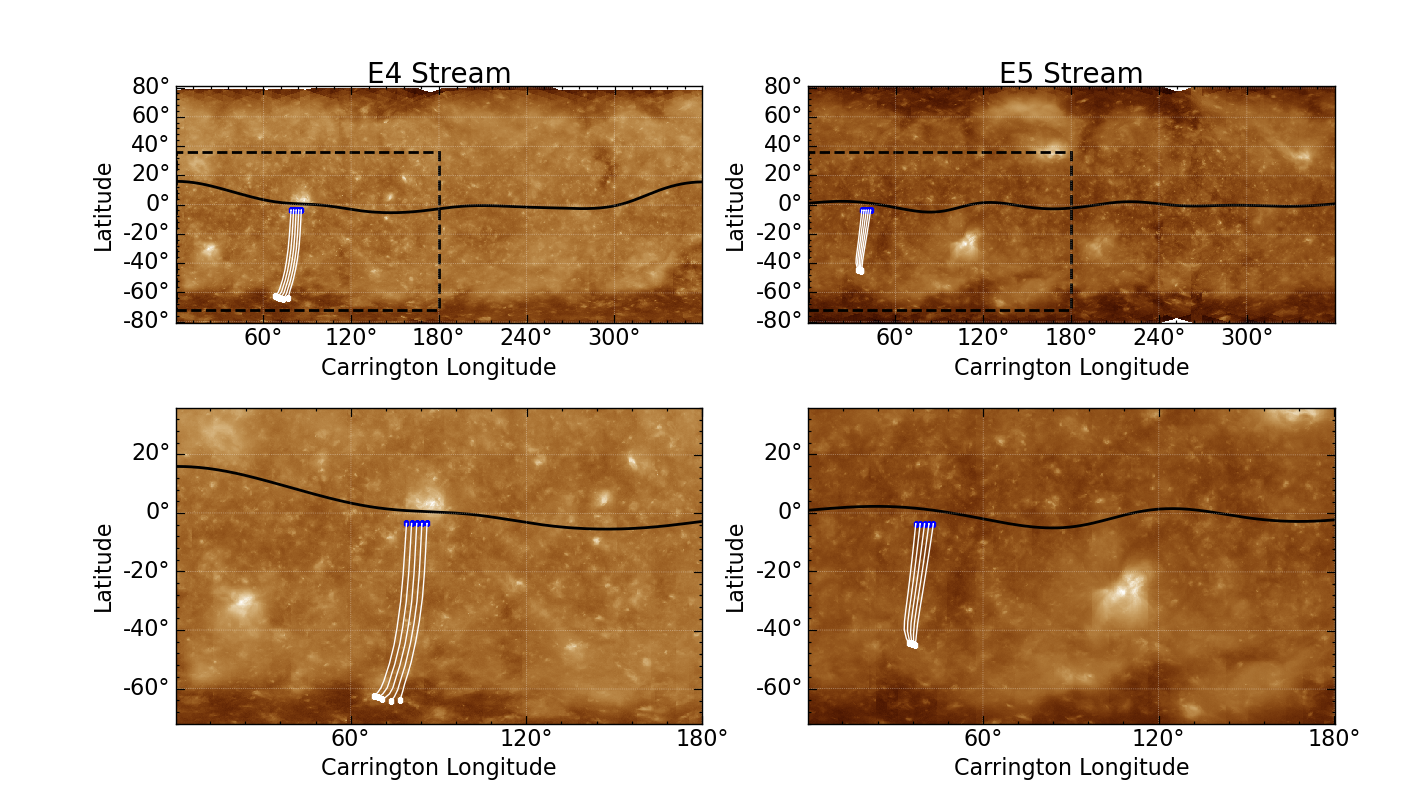}
\caption[]{\small Potential field source surface mapping for the encounter 4 (left) and encounter 5 (right) streams. The top row shows the whole Sun and the bottom row shows a close up of the area of interest for each stream (indicated by the black dashed box in the top row). The white lines show the mapping of the magnetic field lines from the solar surface (white points) to the source surface (blue points). The solid black line shows the heliospheric current sheet. Both streams map to the edge of the Southern polar coronal hole.}
\label{pfss_map}
\end{figure*}


\bibliographystyle{mnras}
\bibliography{references}

\begin{thebibliography}{}
\makeatletter
\relax
\def\mn@urlcharsother{\let\do\@makeother \do\$\do\&\do\#\do\^\do\_\do\%\do\~}
\def\mn@doi{\begingroup\mn@urlcharsother \@ifnextchar [ {\mn@doi@}
  {\mn@doi@[]}}
\def\mn@doi@[#1]#2{\def\@tempa{#1}\ifx\@tempa\@empty \href
  {http://dx.doi.org/#2} {doi:#2}\else \href {http://dx.doi.org/#2} {#1}\fi
  \endgroup}
\def\mn@eprint#1#2{\mn@eprint@#1:#2::\@nil}
\def\mn@eprint@arXiv#1{\href {http://arxiv.org/abs/#1} {{\tt arXiv:#1}}}
\def\mn@eprint@dblp#1{\href {http://dblp.uni-trier.de/rec/bibtex/#1.xml}
  {dblp:#1}}
\def\mn@eprint@#1:#2:#3:#4\@nil{\def\@tempa {#1}\def\@tempb {#2}\def\@tempc
  {#3}\ifx \@tempc \@empty \let \@tempc \@tempb \let \@tempb \@tempa \fi \ifx
  \@tempb \@empty \def\@tempb {arXiv}\fi \@ifundefined
  {mn@eprint@\@tempb}{\@tempb:\@tempc}{\expandafter \expandafter \csname
  mn@eprint@\@tempb\endcsname \expandafter{\@tempc}}}

\bibitem[\protect\citeauthoryear{{Abbo} et~al.,}{{Abbo}
  et~al.}{2016}]{Abbo_2016}
{Abbo} L.,  et~al., 2016, \mn@doi [\ssr] {10.1007/s11214-016-0264-1}, \href
  {https://ui.adsabs.harvard.edu/abs/2016SSRv..201...55A} {201, 55}

\bibitem[\protect\citeauthoryear{{Alterman}, {Kasper}, {Stevens}  \&
  {Koval}}{{Alterman} et~al.}{2018}]{Alterman_2018}
{Alterman} B.~L.,  {Kasper} J.~C.,  {Stevens} M.~L.,   {Koval} A.,  2018,
  \mn@doi [\apj] {10.3847/1538-4357/aad23f}, \href
  {https://ui.adsabs.harvard.edu/abs/2018ApJ...864..112A} {864, 112}

\bibitem[\protect\citeauthoryear{{Alterman}, {Kasper}, {Leamon}  \&
  {McIntosh}}{{Alterman} et~al.}{2021}]{Alterman_2021}
{Alterman} B.~L.,  {Kasper} J.~C.,  {Leamon} R.~J.,   {McIntosh} S.~W.,  2021,
  \mn@doi [\solphys] {10.1007/s11207-021-01801-9}, \href
  {https://ui.adsabs.harvard.edu/abs/2021SoPh..296...67A} {296, 67}

\bibitem[\protect\citeauthoryear{{Altschuler} \& {Newkirk}}{{Altschuler} \&
  {Newkirk}}{1969}]{Altschuler_1969}
{Altschuler} M.~D.,  {Newkirk} G.,  1969, \mn@doi [\solphys]
  {10.1007/BF00145734}, \href
  {https://ui.adsabs.harvard.edu/abs/1969SoPh....9..131A} {9, 131}

\bibitem[\protect\citeauthoryear{{Badman} et~al.,}{{Badman}
  et~al.}{2020}]{Badman_2020}
{Badman} S.~T.,  et~al., 2020, \mn@doi [\apjs] {10.3847/1538-4365/ab4da7},
  \href {https://ui.adsabs.harvard.edu/abs/2020ApJS..246...23B} {246, 23}

\bibitem[\protect\citeauthoryear{{Bale} et~al.,}{{Bale}
  et~al.}{2016}]{Bale_2016}
{Bale} S.~D.,  et~al., 2016, \mn@doi [\ssr] {10.1007/s11214-016-0244-5}, \href
  {https://ui.adsabs.harvard.edu/abs/2016SSRv..204...49B} {204, 49}

\bibitem[\protect\citeauthoryear{{Bale} et~al.,}{{Bale}
  et~al.}{2019}]{Bale_Nature_2019}
{Bale} S.~D.,  et~al., 2019, \mn@doi [\nat] {10.1038/s41586-019-1818-7}, \href
  {https://ui.adsabs.harvard.edu/abs/2019Natur.576..237B} {576, 237}

\bibitem[\protect\citeauthoryear{{Bame}, {Asbridge}, {Feldman}  \&
  {Gosling}}{{Bame} et~al.}{1977}]{Bame_1977}
{Bame} S.~J.,  {Asbridge} J.~R.,  {Feldman} W.~C.,   {Gosling} J.~T.,  1977,
  \mn@doi [\jgr] {10.1029/JA082i010p01487}, \href
  {https://ui.adsabs.harvard.edu/abs/1977JGR....82.1487B} {82, 1487}

\bibitem[\protect\citeauthoryear{{Berger}, {Wimmer-Schweingruber}  \&
  {Gloeckler}}{{Berger} et~al.}{2011}]{Berger_2011}
{Berger} L.,  {Wimmer-Schweingruber} R.~F.,   {Gloeckler} G.,  2011, \mn@doi
  [\prl] {10.1103/PhysRevLett.106.151103}, \href
  {https://ui.adsabs.harvard.edu/abs/2011PhRvL.106o1103B} {106, 151103}

\bibitem[\protect\citeauthoryear{{Ber{\v{c}}i{\v{c}}}
  et~al.,}{{Ber{\v{c}}i{\v{c}}} et~al.}{2020}]{Bercic_2020}
{Ber{\v{c}}i{\v{c}}} L.,  et~al., 2020, \mn@doi [\apj]
  {10.3847/1538-4357/ab7b7a}, \href
  {https://ui.adsabs.harvard.edu/abs/2020ApJ...892...88B} {892, 88}

\bibitem[\protect\citeauthoryear{{Bourouaine} \& {Chandran}}{{Bourouaine} \&
  {Chandran}}{2013}]{Bourouaine_2013}
{Bourouaine} S.,  {Chandran} B. D.~G.,  2013, \mn@doi [\apj]
  {10.1088/0004-637X/774/2/96}, \href
  {https://ui.adsabs.harvard.edu/abs/2013ApJ...774...96B} {774, 96}

\bibitem[\protect\citeauthoryear{{Breen}, {Riley}, {Lazarus}, {Canals},
  {Fallows}, {Linker}  \& {Mikic}}{{Breen} et~al.}{2002}]{Breen_2002}
{Breen} A.~R.,  {Riley} P.,  {Lazarus} A.~J.,  {Canals} A.,  {Fallows} R.~A.,
  {Linker} J.,   {Mikic} Z.,  2002, \mn@doi [Annales Geophysicae]
  {10.5194/angeo-20-1291-2002}, \href
  {https://ui.adsabs.harvard.edu/abs/2002AnGeo..20.1291B} {20, 1291}

\bibitem[\protect\citeauthoryear{{Burlaga} \& {Ogilvie}}{{Burlaga} \&
  {Ogilvie}}{1970}]{Burlaga_1970}
{Burlaga} L.~F.,  {Ogilvie} K.~W.,  1970, \mn@doi [\apj] {10.1086/150340},
  \href {https://ui.adsabs.harvard.edu/abs/1970ApJ...159..659B} {159, 659}

\bibitem[\protect\citeauthoryear{{Byhring}}{{Byhring}}{2011}]{Byhring_2011}
{Byhring} H.~S.,  2011, \mn@doi [\apj] {10.1088/0004-637X/738/2/172}, \href
  {https://ui.adsabs.harvard.edu/abs/2011ApJ...738..172B} {738, 172}

\bibitem[\protect\citeauthoryear{{Camporeale}, {Car{\`e}}  \&
  {Borovsky}}{{Camporeale} et~al.}{2017}]{Camporeale_2017}
{Camporeale} E.,  {Car{\`e}} A.,   {Borovsky} J.~E.,  2017, \mn@doi [Journal of
  Geophysical Research (Space Physics)] {10.1002/2017JA024383}, \href
  {https://ui.adsabs.harvard.edu/abs/2017JGRA..12210910C} {122, 10,910}

\bibitem[\protect\citeauthoryear{{Chandran}}{{Chandran}}{2010}]{Chandran_2010}
{Chandran} B. D.~G.,  2010, \mn@doi [\apj] {10.1088/0004-637X/720/1/548}, \href
  {https://ui.adsabs.harvard.edu/abs/2010ApJ...720..548C} {720, 548}

\bibitem[\protect\citeauthoryear{{Chen} et~al.,}{{Chen}
  et~al.}{2020}]{Chen_2020}
{Chen} C.~H.~K.,  et~al., 2020, \mn@doi [\apjs] {10.3847/1538-4365/ab60a3},
  \href {https://ui.adsabs.harvard.edu/abs/2020ApJS..246...53C} {246, 53}

\bibitem[\protect\citeauthoryear{{Chen} et~al.,}{{Chen}
  et~al.}{2021}]{Chen_2021}
{Chen} C. H.~K.,  et~al., 2021, \mn@doi [A\&A] {10.1051/0004-6361/202039872},
  (in press)

\bibitem[\protect\citeauthoryear{{Cranmer}, {van Ballegooijen}  \&
  {Edgar}}{{Cranmer} et~al.}{2007}]{Cranmer_2007}
{Cranmer} S.~R.,  {van Ballegooijen} A.~A.,   {Edgar} R.~J.,  2007, \mn@doi
  [\apjs] {10.1086/518001}, \href
  {https://ui.adsabs.harvard.edu/abs/2007ApJS..171..520C} {171, 520}

\bibitem[\protect\citeauthoryear{{D'Amicis} \& {Bruno}}{{D'Amicis} \&
  {Bruno}}{2015}]{Damicis_Bruno_2015}
{D'Amicis} R.,  {Bruno} R.,  2015, \mn@doi [\apj] {10.1088/0004-637X/805/1/84},
  \href {https://ui.adsabs.harvard.edu/abs/2015ApJ...805...84D} {805, 84}

\bibitem[\protect\citeauthoryear{{D'Amicis}, {Matteini}  \& {Bruno}}{{D'Amicis}
  et~al.}{2019}]{Damicis_2019}
{D'Amicis} R.,  {Matteini} L.,   {Bruno} R.,  2019, \mn@doi [\mnras]
  {10.1093/mnras/sty3329}, \href
  {https://ui.adsabs.harvard.edu/abs/2019MNRAS.483.4665D} {483, 4665}

\bibitem[\protect\citeauthoryear{{Dudok de Wit} et~al.,}{{Dudok de Wit}
  et~al.}{2020}]{DudokdeWit_2020}
{Dudok de Wit} T.,  et~al., 2020, \mn@doi [\apjs] {10.3847/1538-4365/ab5853},
  \href {https://ui.adsabs.harvard.edu/abs/2020ApJS..246...39D} {246, 39}

\bibitem[\protect\citeauthoryear{{Einaudi}, {Boncinelli}, {Dahlburg}  \&
  {Karpen}}{{Einaudi} et~al.}{1999}]{Einaudi_1999}
{Einaudi} G.,  {Boncinelli} P.,  {Dahlburg} R.~B.,   {Karpen} J.~T.,  1999,
  \mn@doi [\jgr] {10.1029/98JA02394}, \href
  {https://ui.adsabs.harvard.edu/abs/1999JGR...104..521E} {104, 521}

\bibitem[\protect\citeauthoryear{{Feldman}, {Asbridge}, {Bame}  \&
  {Gosling}}{{Feldman} et~al.}{1978}]{feldman_1978}
{Feldman} W.~C.,  {Asbridge} J.~R.,  {Bame} S.~J.,   {Gosling} J.~T.,  1978,
  \mn@doi [\jgr] {10.1029/JA083iA05p02177}, \href
  {https://ui.adsabs.harvard.edu/abs/1978JGR....83.2177F} {83, 2177}

\bibitem[\protect\citeauthoryear{{Feldman}, {Widing}  \& {Warren}}{{Feldman}
  et~al.}{1999}]{Feldman_1999}
{Feldman} U.,  {Widing} K.~G.,   {Warren} H.~P.,  1999, \mn@doi [\apj]
  {10.1086/307682}, \href
  {https://ui.adsabs.harvard.edu/abs/1999ApJ...522.1133F} {522, 1133}

\bibitem[\protect\citeauthoryear{{Finley} et~al.,}{{Finley}
  et~al.}{2020}]{Finley_2020}
{Finley} A.~J.,  et~al., 2020, \mn@doi [A\&A] {10.1051/0004-6361/202039288},
  (in press)

\bibitem[\protect\citeauthoryear{{Fisk}}{{Fisk}}{2003}]{Fisk_2003}
{Fisk} L.~A.,  2003, \mn@doi [Journal of Geophysical Research (Space Physics)]
  {10.1029/2002JA009284}, \href
  {https://ui.adsabs.harvard.edu/abs/2003JGRA..108.1157F} {108, 1157}

\bibitem[\protect\citeauthoryear{{Fisk} \& {Schwadron}}{{Fisk} \&
  {Schwadron}}{2001}]{Fisk_2001}
{Fisk} L.~A.,  {Schwadron} N.~A.,  2001, \mn@doi [\apj] {10.1086/322503}, \href
  {https://ui.adsabs.harvard.edu/abs/2001ApJ...560..425F} {560, 425}

\bibitem[\protect\citeauthoryear{{Fisk}, {Schwadron}  \& {Zurbuchen}}{{Fisk}
  et~al.}{1999}]{Fisk_1999}
{Fisk} L.~A.,  {Schwadron} N.~A.,   {Zurbuchen} T.~H.,  1999, \mn@doi [\jgr]
  {10.1029/1999JA900256}, \href
  {https://ui.adsabs.harvard.edu/abs/1999JGR...10419765F} {104, 19765}

\bibitem[\protect\citeauthoryear{{Fox} et~al.,}{{Fox} et~al.}{2016}]{Fox_2016}
{Fox} N.~J.,  et~al., 2016, \mn@doi [\ssr] {10.1007/s11214-015-0211-6}, \href
  {https://ui.adsabs.harvard.edu/abs/2016SSRv..204....7F} {204, 7}

\bibitem[\protect\citeauthoryear{{Fu}, {Madjarska}, {Li}, {Xia}  \&
  {Huang}}{{Fu} et~al.}{2018}]{Fu_2018}
{Fu} H.,  {Madjarska} M.~S.,  {Li} B.,  {Xia} L.,   {Huang} Z.,  2018, \mn@doi
  [\mnras] {10.1093/mnras/sty1211}, \href
  {https://ui.adsabs.harvard.edu/abs/2018MNRAS.478.1884F} {478, 1884}

\bibitem[\protect\citeauthoryear{{Geiss}, {Hirt}  \& {Leutwyler}}{{Geiss}
  et~al.}{1970}]{Geiss_1970}
{Geiss} J.,  {Hirt} P.,   {Leutwyler} H.,  1970, \mn@doi [\solphys]
  {10.1007/BF00148028}, \href
  {https://ui.adsabs.harvard.edu/abs/1970SoPh...12..458G} {12, 458}

\bibitem[\protect\citeauthoryear{{Gloeckler}, {Zurbuchen}  \&
  {Geiss}}{{Gloeckler} et~al.}{2003}]{Gloeckler_2003}
{Gloeckler} G.,  {Zurbuchen} T.~H.,   {Geiss} J.,  2003, \mn@doi [Journal of
  Geophysical Research (Space Physics)] {10.1029/2002JA009286}, \href
  {https://ui.adsabs.harvard.edu/abs/2003JGRA..108.1158G} {108, 1158}

\bibitem[\protect\citeauthoryear{{Gosling}, {Asbridge}, {Bame}  \&
  {Feldman}}{{Gosling} et~al.}{1978}]{Gosling_1978}
{Gosling} J.~T.,  {Asbridge} J.~R.,  {Bame} S.~J.,   {Feldman} W.~C.,  1978,
  \mn@doi [\jgr] {10.1029/JA083iA04p01401}, \href
  {https://ui.adsabs.harvard.edu/abs/1978JGR....83.1401G} {83, 1401}

\bibitem[\protect\citeauthoryear{{Harra}, {Sakao}, {Mandrini}, {Hara}, {Imada},
  {Young}, {van Driel-Gesztelyi}  \& {Baker}}{{Harra}
  et~al.}{2008}]{Harra_2008}
{Harra} L.~K.,  {Sakao} T.,  {Mandrini} C.~H.,  {Hara} H.,  {Imada} S.,
  {Young} P.~R.,  {van Driel-Gesztelyi} L.,   {Baker} D.,  2008, \mn@doi
  [\apjl] {10.1086/587485}, \href
  {https://ui.adsabs.harvard.edu/abs/2008ApJ...676L.147H} {676, L147}

\bibitem[\protect\citeauthoryear{{Hellinger} \&
  {Tr{\'a}vn{\'\i}{\v{c}}ek}}{{Hellinger} \&
  {Tr{\'a}vn{\'\i}{\v{c}}ek}}{2013}]{Hellinger_2013}
{Hellinger} P.,  {Tr{\'a}vn{\'\i}{\v{c}}ek} P.~M.,  2013, \mn@doi [Journal of
  Geophysical Research (Space Physics)] {10.1002/jgra.50540}, \href
  {https://ui.adsabs.harvard.edu/abs/2013JGRA..118.5421H} {118, 5421}

\bibitem[\protect\citeauthoryear{{Hellinger}, {Matteini}, {{\v{S}}tver{\'a}k},
  {Tr{\'a}vn{\'\i}{\v{c}}ek}  \& {Marsch}}{{Hellinger}
  et~al.}{2011}]{Hellinger_2011}
{Hellinger} P.,  {Matteini} L.,  {{\v{S}}tver{\'a}k} {\v{S}}.,
  {Tr{\'a}vn{\'\i}{\v{c}}ek} P.~M.,   {Marsch} E.,  2011, \mn@doi [Journal of
  Geophysical Research (Space Physics)] {10.1029/2011JA016674}, \href
  {https://ui.adsabs.harvard.edu/abs/2011JGRA..116.9105H} {116, A09105}

\bibitem[\protect\citeauthoryear{{Hirshberg}, {Bame}  \& {Robbins}}{{Hirshberg}
  et~al.}{1972a}]{Hirshberg_1972b}
{Hirshberg} J.,  {Bame} S.~J.,   {Robbins} D.~E.,  1972a, \mn@doi [\solphys]
  {10.1007/BF00148109}, \href
  {https://ui.adsabs.harvard.edu/abs/1972SoPh...23..467H} {23, 467}

\bibitem[\protect\citeauthoryear{{Hirshberg}, {Asbridge}  \&
  {Robbins}}{{Hirshberg} et~al.}{1972b}]{Hirshberg_1972}
{Hirshberg} J.,  {Asbridge} J.~R.,   {Robbins} D.~E.,  1972b, \mn@doi [\jgr]
  {10.1029/JA077i019p03583}, \href
  {https://ui.adsabs.harvard.edu/abs/1972JGR....77.3583H} {77, 3583}

\bibitem[\protect\citeauthoryear{{Horbury}, {Matteini}  \& {Stansby}}{{Horbury}
  et~al.}{2018}]{Horbury_2018}
{Horbury} T.~S.,  {Matteini} L.,   {Stansby} D.,  2018, \mn@doi [\mnras]
  {10.1093/mnras/sty953}, \href
  {https://ui.adsabs.harvard.edu/abs/2018MNRAS.478.1980H} {478, 1980}

\bibitem[\protect\citeauthoryear{{Horbury} et~al.,}{{Horbury}
  et~al.}{2020}]{Horbury_2020}
{Horbury} T.~S.,  et~al., 2020, \mn@doi [\apjs] {10.3847/1538-4365/ab5b15},
  \href {https://ui.adsabs.harvard.edu/abs/2020ApJS..246...45H} {246, 45}

\bibitem[\protect\citeauthoryear{{Huang} et~al.,}{{Huang}
  et~al.}{2020}]{Huang_2020_b}
{Huang} J.,  et~al., 2020, arXiv e-prints, \href
  {https://ui.adsabs.harvard.edu/abs/2020arXiv200512372H} {p. arXiv:2005.12372}

\bibitem[\protect\citeauthoryear{{Jockers}}{{Jockers}}{1970}]{Jockers_1970}
{Jockers} K.,  1970, \aap, \href
  {https://ui.adsabs.harvard.edu/abs/1970A&A.....6..219J} {6, 219}

\bibitem[\protect\citeauthoryear{{Kasper}, {Stevens}, {Lazarus}, {Steinberg}
  \& {Ogilvie}}{{Kasper} et~al.}{2007}]{Kasper_2007}
{Kasper} J.~C.,  {Stevens} M.~L.,  {Lazarus} A.~J.,  {Steinberg} J.~T.,
  {Ogilvie} K.~W.,  2007, \mn@doi [\apj] {10.1086/510842}, \href
  {https://ui.adsabs.harvard.edu/abs/2007ApJ...660..901K} {660, 901}

\bibitem[\protect\citeauthoryear{{Kasper}, {Stevens}, {Korreck}, {Maruca},
  {Kiefer}, {Schwadron}  \& {Lepri}}{{Kasper} et~al.}{2012}]{Kasper_2012}
{Kasper} J.~C.,  {Stevens} M.~L.,  {Korreck} K.~E.,  {Maruca} B.~A.,  {Kiefer}
  K.~K.,  {Schwadron} N.~A.,   {Lepri} S.~T.,  2012, \mn@doi [\apj]
  {10.1088/0004-637X/745/2/162}, \href
  {https://ui.adsabs.harvard.edu/abs/2012ApJ...745..162K} {745, 162}

\bibitem[\protect\citeauthoryear{{Kasper} et~al.,}{{Kasper}
  et~al.}{2016}]{Kasper_2016}
{Kasper} J.~C.,  et~al., 2016, \mn@doi [\ssr] {10.1007/s11214-015-0206-3},
  \href {https://ui.adsabs.harvard.edu/abs/2016SSRv..204..131K} {204, 131}

\bibitem[\protect\citeauthoryear{{Kasper} et~al.,}{{Kasper}
  et~al.}{2019}]{Kasper_nature_2019}
{Kasper} J.~C.,  et~al., 2019, \mn@doi [\nat] {10.1038/s41586-019-1813-z},
  \href {https://ui.adsabs.harvard.edu/abs/2019Natur.576..228K} {576, 228}

\bibitem[\protect\citeauthoryear{{Krieger}, {Timothy}  \& {Roelof}}{{Krieger}
  et~al.}{1973}]{Krieger_1973}
{Krieger} A.~S.,  {Timothy} A.~F.,   {Roelof} E.~C.,  1973, \mn@doi [\solphys]
  {10.1007/BF00150828}, \href
  {https://ui.adsabs.harvard.edu/abs/1973SoPh...29..505K} {29, 505}

\bibitem[\protect\citeauthoryear{{Laker} et~al.,}{{Laker}
  et~al.}{2020}]{Laker_2020}
{Laker} R.,  et~al., 2020, \mn@doi [A\&A] {10.1051/0004-6361/202039354}, (in
  press)

\bibitem[\protect\citeauthoryear{{Laker} et~al.,}{{Laker}
  et~al.}{2021}]{Laker_2021}
{Laker} R.,  et~al., 2021, arXiv e-prints, \href
  {https://ui.adsabs.harvard.edu/abs/2021arXiv210300230L} {p. arXiv:2103.00230}

\bibitem[\protect\citeauthoryear{{Laming}}{{Laming}}{2004}]{Laming_2004}
{Laming} J.~M.,  2004, \mn@doi [\apj] {10.1086/423780}, \href
  {https://ui.adsabs.harvard.edu/abs/2004ApJ...614.1063L} {614, 1063}

\bibitem[\protect\citeauthoryear{{Lavraud} et~al.,}{{Lavraud}
  et~al.}{2020}]{Lavraud_2020}
{Lavraud} B.,  et~al., 2020, \mn@doi [\apjl] {10.3847/2041-8213/ab8d2d}, \href
  {https://ui.adsabs.harvard.edu/abs/2020ApJ...894L..19L} {894, L19}

\bibitem[\protect\citeauthoryear{{Lemaire} \& {Scherer}}{{Lemaire} \&
  {Scherer}}{1970}]{Lemaire_1970}
{Lemaire} J.,  {Scherer} M.,  1970, \mn@doi [\planss]
  {10.1016/0032-0633(70)90070-X}, \href
  {https://ui.adsabs.harvard.edu/abs/1970P&SS...18..103L} {18, 103}

\bibitem[\protect\citeauthoryear{{Lemaire} \& {Scherer}}{{Lemaire} \&
  {Scherer}}{1971}]{Lemaire_1971}
{Lemaire} J.,  {Scherer} M.,  1971, \mn@doi [\jgr] {10.1029/JA076i031p07479},
  \href {https://ui.adsabs.harvard.edu/abs/1971JGR....76.7479L} {76, 7479}

\bibitem[\protect\citeauthoryear{{Li}, {Habbal}, {Hollweg}  \& {Esser}}{{Li}
  et~al.}{1999}]{Li_1999}
{Li} X.,  {Habbal} S.~R.,  {Hollweg} J.~V.,   {Esser} R.,  1999, \mn@doi [\jgr]
  {10.1029/1998JA900126}, \href
  {https://ui.adsabs.harvard.edu/abs/1999JGR...104.2521L} {104, 2521}

\bibitem[\protect\citeauthoryear{{Li}, {Wang}, {Tu}  \& {Xu}}{{Li}
  et~al.}{2020}]{Li_2020}
{Li} H.,  {Wang} C.,  {Tu} C.,   {Xu} F.,  2020, \mn@doi [Earth and Space
  Science] {10.1029/2019EA000997}, \href
  {https://ui.adsabs.harvard.edu/abs/2020E&SS....700997L} {7, e00997}

\bibitem[\protect\citeauthoryear{{Lopez} \& {Freeman}}{{Lopez} \&
  {Freeman}}{1986}]{Lopez_1986}
{Lopez} R.~E.,  {Freeman} J.~W.,  1986, \mn@doi [\jgr]
  {10.1029/JA091iA02p01701}, \href
  {https://ui.adsabs.harvard.edu/abs/1986JGR....91.1701L} {91, 1701}

\bibitem[\protect\citeauthoryear{{Maksimovic}, {Pierrard}  \&
  {Lemaire}}{{Maksimovic} et~al.}{1997}]{Maksimovic_1997}
{Maksimovic} M.,  {Pierrard} V.,   {Lemaire} J.~F.,  1997, \aap, \href
  {https://ui.adsabs.harvard.edu/abs/1997A&A...324..725M} {324, 725}

\bibitem[\protect\citeauthoryear{{Marsch}, {Rosenbauer}, {Schwenn},
  {Muehlhaeuser}  \& {Neubauer}}{{Marsch} et~al.}{1982a}]{Marsch_1982}
{Marsch} E.,  {Rosenbauer} H.,  {Schwenn} R.,  {Muehlhaeuser} K.~H.,
  {Neubauer} F.~M.,  1982a, \mn@doi [\jgr] {10.1029/JA087iA01p00035}, \href
  {https://ui.adsabs.harvard.edu/abs/1982JGR....87...35M} {87, 35}

\bibitem[\protect\citeauthoryear{{Marsch}, {Schwenn}, {Rosenbauer},
  {Muehlhaeuser}, {Pilipp}  \& {Neubauer}}{{Marsch}
  et~al.}{1982b}]{Marsch_1982_protons}
{Marsch} E.,  {Schwenn} R.,  {Rosenbauer} H.,  {Muehlhaeuser} K.~H.,  {Pilipp}
  W.,   {Neubauer} F.~M.,  1982b, \mn@doi [\jgr] {10.1029/JA087iA01p00052},
  \href {https://ui.adsabs.harvard.edu/abs/1982JGR....87...52M} {87, 52}

\bibitem[\protect\citeauthoryear{{Maruca}, {Kasper}  \& {Gary}}{{Maruca}
  et~al.}{2012}]{Maruca_2012}
{Maruca} B.~A.,  {Kasper} J.~C.,   {Gary} S.~P.,  2012, \mn@doi [\apj]
  {10.1088/0004-637X/748/2/137}, \href
  {https://ui.adsabs.harvard.edu/abs/2012ApJ...748..137M} {748, 137}

\bibitem[\protect\citeauthoryear{{Matteini}, {Hellinger}, {Landi},
  {Tr{\'a}vn{\'\i}{\v{c}}ek}  \& {Velli}}{{Matteini}
  et~al.}{2012}]{Matteini_2012}
{Matteini} L.,  {Hellinger} P.,  {Landi} S.,  {Tr{\'a}vn{\'\i}{\v{c}}ek} P.~M.,
    {Velli} M.,  2012, \mn@doi [\ssr] {10.1007/s11214-011-9774-z}, \href
  {https://ui.adsabs.harvard.edu/abs/2012SSRv..172..373M} {172, 373}

\bibitem[\protect\citeauthoryear{{Matteini}, {Horbury}, {Neugebauer}  \&
  {Goldstein}}{{Matteini} et~al.}{2014}]{Matteini_2014}
{Matteini} L.,  {Horbury} T.~S.,  {Neugebauer} M.,   {Goldstein} B.~E.,  2014,
  \mn@doi [\grl] {10.1002/2013GL058482}, \href
  {https://ui.adsabs.harvard.edu/abs/2014GeoRL..41..259M} {41, 259}

\bibitem[\protect\citeauthoryear{{Matteini}, {Horbury}, {Pantellini}, {Velli}
  \& {Schwartz}}{{Matteini} et~al.}{2015a}]{Matteini_2015_alfvenic}
{Matteini} L.,  {Horbury} T.~S.,  {Pantellini} F.,  {Velli} M.,   {Schwartz}
  S.~J.,  2015a, \mn@doi [\apj] {10.1088/0004-637X/802/1/11}, \href
  {https://ui.adsabs.harvard.edu/abs/2015ApJ...802...11M} {802, 11}

\bibitem[\protect\citeauthoryear{{Matteini}, {Hellinger}, {Schwartz}  \&
  {Landi}}{{Matteini} et~al.}{2015b}]{Matteini_2015}
{Matteini} L.,  {Hellinger} P.,  {Schwartz} S.~J.,   {Landi} S.,  2015b,
  \mn@doi [\apj] {10.1088/0004-637X/812/1/13}, \href
  {https://ui.adsabs.harvard.edu/abs/2015ApJ...812...13M} {812, 13}

\bibitem[\protect\citeauthoryear{{M{\"u}ller}, {Marsden}, {St. Cyr}  \&
  {Gilbert}}{{M{\"u}ller} et~al.}{2013}]{Muller_2013}
{M{\"u}ller} D.,  {Marsden} R.~G.,  {St. Cyr} O.~C.,   {Gilbert} H.~R.,  2013,
  \mn@doi [\solphys] {10.1007/s11207-012-0085-7}, \href
  {https://ui.adsabs.harvard.edu/abs/2013SoPh..285...25M} {285, 25}

\bibitem[\protect\citeauthoryear{{Neugebauer}}{{Neugebauer}}{1981}]{Neugebauer_1981}
{Neugebauer} M.,  1981, \fcp, \href
  {https://ui.adsabs.harvard.edu/abs/1981FCPh....7..131N} {7, 131}

\bibitem[\protect\citeauthoryear{{Neugebauer} \& {Goldstein}}{{Neugebauer} \&
  {Goldstein}}{2013}]{Neugebauer_2013}
{Neugebauer} M.,  {Goldstein} B.~E.,  2013, in {Zank} G.~P.,  et~al., eds,
  American Institute of Physics Conference Series Vol. 1539, Solar Wind 13. pp
  46--49, \mn@doi{10.1063/1.4810986}

\bibitem[\protect\citeauthoryear{{Neugebauer} et~al.,}{{Neugebauer}
  et~al.}{1998}]{Neugebauer_1998}
{Neugebauer} M.,  et~al., 1998, \mn@doi [\jgr] {10.1029/98JA00798}, \href
  {https://ui.adsabs.harvard.edu/abs/1998JGR...10314587N} {103, 14587}

\bibitem[\protect\citeauthoryear{{Neugebauer}, {Liewer}, {Smith}, {Skoug}  \&
  {Zurbuchen}}{{Neugebauer} et~al.}{2002}]{Neugebauer_2002}
{Neugebauer} M.,  {Liewer} P.~C.,  {Smith} E.~J.,  {Skoug} R.~M.,   {Zurbuchen}
  T.~H.,  2002, \mn@doi [Journal of Geophysical Research (Space Physics)]
  {10.1029/2001JA000306}, \href
  {https://ui.adsabs.harvard.edu/abs/2002JGRA..107.1488N} {107, 1488}

\bibitem[\protect\citeauthoryear{{Nolte} et~al.,}{{Nolte}
  et~al.}{1976}]{Nolte_1976}
{Nolte} J.~T.,  et~al., 1976, \mn@doi [\solphys] {10.1007/BF00149859}, \href
  {https://ui.adsabs.harvard.edu/abs/1976SoPh...46..303N} {46, 303}

\bibitem[\protect\citeauthoryear{{O'Kane} et~al.,}{{O'Kane}
  et~al.}{2021}]{Kane_2021}
{O'Kane} J.,  et~al., 2021, \mn@doi [A\&A] {10.1051/0004-6361/202140622}, (in
  press)

\bibitem[\protect\citeauthoryear{{Ogilvie} \& {Wilkerson}}{{Ogilvie} \&
  {Wilkerson}}{1969}]{Ogilvie_1969}
{Ogilvie} K.~W.,  {Wilkerson} T.~D.,  1969, \mn@doi [\solphys]
  {10.1007/BF00155391}, \href
  {https://ui.adsabs.harvard.edu/abs/1969SoPh....8..435O} {8, 435}

\bibitem[\protect\citeauthoryear{{Perrone}, {Stansby}, {Horbury}  \&
  {Matteini}}{{Perrone} et~al.}{2019}]{Perrone_2019_thermo}
{Perrone} D.,  {Stansby} D.,  {Horbury} T.~S.,   {Matteini} L.,  2019, \mn@doi
  [\mnras] {10.1093/mnras/stz1877}, \href
  {https://ui.adsabs.harvard.edu/abs/2019MNRAS.488.2380P} {488, 2380}

\bibitem[\protect\citeauthoryear{{Perrone}, {D'Amicis}, {De Marco}, {Matteini},
  {Stansby}, {Bruno}  \& {Horbury}}{{Perrone} et~al.}{2020}]{Perrone_2020}
{Perrone} D.,  {D'Amicis} R.,  {De Marco} R.,  {Matteini} L.,  {Stansby} D.,
  {Bruno} R.,   {Horbury} T.~S.,  2020, \mn@doi [\aap]
  {10.1051/0004-6361/201937064}, \href
  {https://ui.adsabs.harvard.edu/abs/2020A&A...633A.166P} {633, A166}

\bibitem[\protect\citeauthoryear{{Rakowski} \& {Laming}}{{Rakowski} \&
  {Laming}}{2012}]{Rakowski_2012}
{Rakowski} C.~E.,  {Laming} J.~M.,  2012, \mn@doi [\apj]
  {10.1088/0004-637X/754/1/65}, \href
  {https://ui.adsabs.harvard.edu/abs/2012ApJ...754...65R} {754, 65}

\bibitem[\protect\citeauthoryear{{Robbins}, {Hundhausen}  \& {Bame}}{{Robbins}
  et~al.}{1970}]{Robbins_1970}
{Robbins} D.~E.,  {Hundhausen} A.~J.,   {Bame} S.~J.,  1970, \mn@doi [\jgr]
  {10.1029/JA075i007p01178}, \href
  {https://ui.adsabs.harvard.edu/abs/1970JGR....75.1178R} {75, 1178}

\bibitem[\protect\citeauthoryear{{Sanchez-Diaz}, {Rouillard}, {Lavraud},
  {Segura}, {Tao}, {Pinto}, {Sheeley}  \& {Plotnikov}}{{Sanchez-Diaz}
  et~al.}{2016}]{Sanchez_Diaz_2016}
{Sanchez-Diaz} E.,  {Rouillard} A.~P.,  {Lavraud} B.,  {Segura} K.,  {Tao} C.,
  {Pinto} R.,  {Sheeley} N.~R.,   {Plotnikov} I.,  2016, \mn@doi [Journal of
  Geophysical Research (Space Physics)] {10.1002/2016JA022433}, \href
  {https://ui.adsabs.harvard.edu/abs/2016JGRA..121.2830S} {121, 2830}

\bibitem[\protect\citeauthoryear{{Schatten}, {Wilcox}  \& {Ness}}{{Schatten}
  et~al.}{1969}]{Schatten_1969}
{Schatten} K.~H.,  {Wilcox} J.~M.,   {Ness} N.~F.,  1969, \mn@doi [\solphys]
  {10.1007/BF00146478}, \href
  {https://ui.adsabs.harvard.edu/abs/1969SoPh....6..442S} {6, 442}

\bibitem[\protect\citeauthoryear{{Sheeley}, {Harvey}  \& {Feldman}}{{Sheeley}
  et~al.}{1976}]{Sheeley_1976}
{Sheeley} N.~R. J.,  {Harvey} J.~W.,   {Feldman} W.~C.,  1976, \mn@doi
  [\solphys] {10.1007/BF00162451}, \href
  {https://ui.adsabs.harvard.edu/abs/1976SoPh...49..271S} {49, 271}

\bibitem[\protect\citeauthoryear{{Shoda}, {Chandran}  \& {Cranmer}}{{Shoda}
  et~al.}{2021}]{Shoda_2021}
{Shoda} M.,  {Chandran} B. D.~G.,   {Cranmer} S.~R.,  2021, arXiv e-prints,
  \href {https://ui.adsabs.harvard.edu/abs/2021arXiv210109529S} {p.
  arXiv:2101.09529}

\bibitem[\protect\citeauthoryear{{Smith}, {Matthaeus}, {Zank}, {Ness},
  {Oughton}  \& {Richardson}}{{Smith} et~al.}{2001}]{Smith_2001}
{Smith} C.~W.,  {Matthaeus} W.~H.,  {Zank} G.~P.,  {Ness} N.~F.,  {Oughton} S.,
    {Richardson} J.~D.,  2001, \mn@doi [\jgr] {10.1029/2000JA000366}, \href
  {https://ui.adsabs.harvard.edu/abs/2001JGR...106.8253S} {106, 8253}

\bibitem[\protect\citeauthoryear{{Squire}, {Chandran}  \& {Meyrand}}{{Squire}
  et~al.}{2020}]{Squire_2020}
{Squire} J.,  {Chandran} B.~D.~G.,   {Meyrand} R.,  2020, \mn@doi [\apjl]
  {10.3847/2041-8213/ab74e1}, \href
  {https://ui.adsabs.harvard.edu/abs/2020ApJ...891L...2S} {891, L2}

\bibitem[\protect\citeauthoryear{{Stansby}, {Salem}, {Matteini}  \&
  {Horbury}}{{Stansby} et~al.}{2018}]{Stansby_2018_helios_fits}
{Stansby} D.,  {Salem} C.,  {Matteini} L.,   {Horbury} T.,  2018, \mn@doi
  [\solphys] {10.1007/s11207-018-1377-3}, \href
  {https://ui.adsabs.harvard.edu/abs/2018SoPh..293..155S} {293, 155}

\bibitem[\protect\citeauthoryear{{Stansby}, {Horbury}  \& {Matteini}}{{Stansby}
  et~al.}{2019a}]{Stansby_2019}
{Stansby} D.,  {Horbury} T.~S.,   {Matteini} L.,  2019a, \mn@doi [\mnras]
  {10.1093/mnras/sty2814}, \href
  {https://ui.adsabs.harvard.edu/abs/2019MNRAS.482.1706S} {482, 1706}

\bibitem[\protect\citeauthoryear{{Stansby}, {Perrone}, {Matteini}, {Horbury}
  \& {Salem}}{{Stansby} et~al.}{2019b}]{Stansby_2019_alphas}
{Stansby} D.,  {Perrone} D.,  {Matteini} L.,  {Horbury} T.~S.,   {Salem} C.~S.,
   2019b, \mn@doi [\aap] {10.1051/0004-6361/201834900}, \href
  {https://ui.adsabs.harvard.edu/abs/2019A&A...623L...2S} {623, L2}

\bibitem[\protect\citeauthoryear{Stansby, Yeates  \& Badman}{Stansby
  et~al.}{2020a}]{Stansby2020_pfsspy}
Stansby D.,  Yeates A.,   Badman S.~T.,  2020a, \mn@doi [Journal of Open Source
  Software] {10.21105/joss.02732}, 5, 2732

\bibitem[\protect\citeauthoryear{{Stansby}, {Matteini}, {Horbury}, {Perrone},
  {D'Amicis}  \& {Ber{\v{c}}i{\v{c}}}}{{Stansby} et~al.}{2020b}]{stansby_2020}
{Stansby} D.,  {Matteini} L.,  {Horbury} T.~S.,  {Perrone} D.,  {D'Amicis} R.,
   {Ber{\v{c}}i{\v{c}}} L.,  2020b, \mn@doi [\mnras] {10.1093/mnras/stz3422},
  \href {https://ui.adsabs.harvard.edu/abs/2020MNRAS.492...39S} {492, 39}

\bibitem[\protect\citeauthoryear{{Sterling} \& {Moore}}{{Sterling} \&
  {Moore}}{2020}]{Sterling_2020}
{Sterling} A.~C.,  {Moore} R.~L.,  2020, \mn@doi [\apjl]
  {10.3847/2041-8213/ab96be}, \href
  {https://ui.adsabs.harvard.edu/abs/2020ApJ...896L..18S} {896, L18}

\bibitem[\protect\citeauthoryear{{Tu}, {Zhou}, {Marsch}, {Xia}, {Zhao}, {Wang}
  \& {Wilhelm}}{{Tu} et~al.}{2005}]{Tu_2005}
{Tu} C.-Y.,  {Zhou} C.,  {Marsch} E.,  {Xia} L.-D.,  {Zhao} L.,  {Wang} J.-X.,
   {Wilhelm} K.,  2005, \mn@doi [Science] {10.1126/science.1109447}, \href
  {https://ui.adsabs.harvard.edu/abs/2005Sci...308..519T} {308, 519}

\bibitem[\protect\citeauthoryear{{Verdini}, {Velli}, {Matthaeus}, {Oughton}  \&
  {Dmitruk}}{{Verdini} et~al.}{2010}]{Verdini_2010}
{Verdini} A.,  {Velli} M.,  {Matthaeus} W.~H.,  {Oughton} S.,   {Dmitruk} P.,
  2010, \mn@doi [\apjl] {10.1088/2041-8205/708/2/L116}, \href
  {https://ui.adsabs.harvard.edu/abs/2010ApJ...708L.116V} {708, L116}

\bibitem[\protect\citeauthoryear{{Verniero} et~al.,}{{Verniero}
  et~al.}{2020}]{Verniero_2020}
{Verniero} J.~L.,  et~al., 2020, \mn@doi [\apjs] {10.3847/1538-4365/ab86af},
  \href {https://ui.adsabs.harvard.edu/abs/2020ApJS..248....5V} {248, 5}

\bibitem[\protect\citeauthoryear{{Verscharen}, {Chandran}, {Bourouaine}  \&
  {Hollweg}}{{Verscharen} et~al.}{2015}]{Verscharen_2015}
{Verscharen} D.,  {Chandran} B. D.~G.,  {Bourouaine} S.,   {Hollweg} J.~V.,
  2015, \mn@doi [\apj] {10.1088/0004-637X/806/2/157}, \href
  {https://ui.adsabs.harvard.edu/abs/2015ApJ...806..157V} {806, 157}

\bibitem[\protect\citeauthoryear{{Verscharen}, {Klein}  \&
  {Maruca}}{{Verscharen} et~al.}{2019}]{Verscharen_2019}
{Verscharen} D.,  {Klein} K.~G.,   {Maruca} B.~A.,  2019, \mn@doi [Living
  Reviews in Solar Physics] {10.1007/s41116-019-0021-0}, \href
  {https://ui.adsabs.harvard.edu/abs/2019LRSP...16....5V} {16, 5}

\bibitem[\protect\citeauthoryear{{Viall} \& {Borovsky}}{{Viall} \&
  {Borovsky}}{2020}]{Viall_2020_review}
{Viall} N.~M.,  {Borovsky} J.~E.,  2020, \mn@doi [Journal of Geophysical
  Research (Space Physics)] {10.1029/2018JA026005}, \href
  {https://ui.adsabs.harvard.edu/abs/2020JGRA..12526005V} {125, e26005}

\bibitem[\protect\citeauthoryear{{Wang} \& {Sheeley}}{{Wang} \&
  {Sheeley}}{1994}]{wang_sheeley_1994}
{Wang} Y.~M.,  {Sheeley} N.~R. J.,  1994, \mn@doi [\jgr] {10.1029/93JA02105},
  \href {https://ui.adsabs.harvard.edu/abs/1994JGR....99.6597W} {99, 6597}

\bibitem[\protect\citeauthoryear{{Whittlesey} et~al.,}{{Whittlesey}
  et~al.}{2020}]{Whittlesey_2020}
{Whittlesey} P.~L.,  et~al., 2020, \mn@doi [\apjs] {10.3847/1538-4365/ab7370},
  \href {https://ui.adsabs.harvard.edu/abs/2020ApJS..246...74W} {246, 74}

\bibitem[\protect\citeauthoryear{{Woodham} et~al.,}{{Woodham}
  et~al.}{2020}]{Woodham_2020}
{Woodham} L.~D.,  et~al., 2020, \mn@doi [A\&A] {10.1051/0004-6361/202039415},
  (in press)

\bibitem[\protect\citeauthoryear{{Woolley} et~al.,}{{Woolley}
  et~al.}{2020}]{Woolley_2020}
{Woolley} T.,  et~al., 2020, \mn@doi [\mnras] {10.1093/mnras/staa2770}, \href
  {https://ui.adsabs.harvard.edu/abs/2020MNRAS.498.5524W} {498, 5524}

\bibitem[\protect\citeauthoryear{{Xu} \& {Borovsky}}{{Xu} \&
  {Borovsky}}{2015}]{xu_borovsky_2015}
{Xu} F.,  {Borovsky} J.~E.,  2015, \mn@doi [Journal of Geophysical Research
  (Space Physics)] {10.1002/2014JA020412}, \href
  {https://ui.adsabs.harvard.edu/abs/2015JGRA..120...70X} {120, 70}

\bibitem[\protect\citeauthoryear{{Zank}, {Nakanotani}, {Zhao}, {Adhikari}  \&
  {Kasper}}{{Zank} et~al.}{2020}]{Zank_2020}
{Zank} G.~P.,  {Nakanotani} M.,  {Zhao} L.~L.,  {Adhikari} L.,   {Kasper} J.,
  2020, \mn@doi [\apj] {10.3847/1538-4357/abb828}, \href
  {https://ui.adsabs.harvard.edu/abs/2020ApJ...903....1Z} {903, 1}

\makeatother
\end{thebibliography}

%\begin{thebibliography}{99}
%\end{thebibliography}
%%%%%%%%%%%%%%%%%%%%%%%%%%%%%%%%%%%%%%%%%%%%%%%%%%

% Don't change these lines
\bsp	% typesetting comment
\label{lastpage}
\end{document}